\documentclass{aastex63}
\usepackage{xcolor}
\usepackage{amsmath}
\usepackage{lipsum}
\usepackage{comment}
\usepackage{makecell}
\usepackage{xstring}

\usepackage{hyperref}
\usepackage{csvsimple}

\usepackage{xcolor}            % colour text
\usepackage{ulem}                          % strikeout

\newcommand{\removed}[1]{\relax}\newcommand{\inserted}[1]{#1}
\newcommand{\skipthis}[1]{\relax}
\received{March 8 2022}
\revised{---}
\accepted{---}
\submitjournal{The Astrophysical Journal Letters}

\shorttitle{When Do Stars Go Boom?}
\shortauthors{Richer et al.}
\graphicspath{{./}{figures/}}
\begin{document}

\title{When Do Stars Go Boom?}

\correspondingauthor{Harvey B. Richer}
\email{richer@astro.ubc.ca}

\author[0000-0001-9002-8178]{Harvey B. Richer}
\affiliation{Department of Physics and Astronomy, University of British Columbia, Vancouver, BC V6T 1Z1, Canada}

\author[0000-0000-0000-0000]{Roger E. Cohen}
\affiliation{Space Telescope Science Institute and Rutgers University}

\author[0000-0000-0000-0000]{Matteo Correnti}
\affiliation{Space Telescope Science Institute, Baltimore, Maryland, 21218 USA}

\author[0000-0002-4770-5388]{Ilaria Caiazzo}
\affiliation{TAPIR, Walter Burke Institute for Theoretical Physics, Mail Code 350-17, Caltech, Pasadena, CA 91125, USA}

\author[0000-0000-0000-0000]{Jeffrey Cummings}
\affiliation{Department of Astronomy, Indiana University, 727 E 3rd Street, Bloomington, In 47405, USA}

\author[0000-0000-0000-0000]{Paul Goudfrooij}
\affiliation{Space Telescope Science Institute, Baltimore, Maryland, 21218 USA}

\author[0000-0000-0000-0000]{Bradley M. S. Hansen}
\affiliation{Mani. L. Bhaumik Institute for Theoretical Physics and Astronomy, University of California, Los Angeles, CA, 90095}

\author[0000-0001-9739-367X]{Jeremy Heyl}
\affiliation{Department of Physics and Astronomy, University of British Columbia, Vancouver, BC V6T 1Z1, Canada}

\author[0000-0000-0000-0000]{Molly Peeples}
\affiliation{Space Telescope Science Institute and Johns Hopkins University, Baltimore, Maryland, USA}

\author[0000-0000-0000-0000]{Jason Kalirai}
\affiliation{Applied Physics Laboratories, Johns Hopkins University, Baltimore, Maryland, USA}

\author[0000-0000-0000-0000]{Elena Sabbi}
\affiliation{Space Telescope Science Institute, Baltimore, Maryland, 21218 USA}

\author[0000-0001-9873-0121]{Pier-Emmanuel Tremblay}
\affiliation{Department of Physics, University of Warwick, Coventry, CV4 7AL, UK}

\author[0000-0000-0000-0000]{Benjamin Williams}
\affiliation{Department of Astronomy, University of Washington, Seattle, Washington 98195-1580, USA}

%\nocollaboration{4}



\begin{abstract}
 The maximum mass of a star that can produce a white dwarf (WD) is an important astrophysical quantity. One of the best approaches to establishing this limit is to search for WDs in young star clusters in which only massive stars have had time to evolve and where the mass of the progenitor can be established from the cooling time of the WD together with the age of the cluster. Searches in young Milky Way clusters have not thus far yielded WD members more massive than about 1.1$~M_{\odot}$, well below the Chandrasekhar mass of $1.38~M_{\odot}$, nor progenitors with masses in excess of about $6~M_{\odot}$. However, the hunt for potentially massive WDs that escaped their cluster environs is yielding interesting candidates. To expand the cluster sample further, we used HST to survey four young and massive star clusters in the Magellanic Clouds for bright WDs that could have evolved from stars as massive as 10$~M_{\odot}$. We located five potential WD candidates in the oldest of the four clusters examined, the first extragalactic single WDs thus far discovered. As these hot WDs are very faint at optical wavelengths, final confirmation will likely have to await spectroscopy with 30-metre class telescopes.

\end{abstract}



\section{Introduction}

The maximum mass of a stable white dwarf (WD) has a widely accepted value of about $1.38~M_{\odot}$ \citep{1987ApJ...322..206N}; the maximum mass of a WD progenitor, however, is much more contentious. Theory suggests this value should be around $8~M_{\odot}$ \citep{1983A&A...121...77W}, but for this limit to hold, the observed type II supernovae (SNe) rate should be much higher than currently observed \citep{2011ApJ...738..154H}. This dearth of observed type II SNe may point to a higher maximum mass, with some theoretical initial mass functions suggesting a maximum progenitor mass closer to $12~M_{\odot}$ \citep[e.g.][]{2003ApJ...598.1076K}. Better constraining this limit is important as it has a profound impact on a number of astrophysical quantities, including, but not limited to, the formation rates of compact objects and the metal enrichment rates of galaxies. 

To probe this limit, we have been searching for massive WDs that are members of young open star clusters \citep{2019ApJ...880...75R,2020ApJ...901L..14C,2021ApJ...912..165R,2021arXiv211009668M,2021arXiv211003837H,2021arXiv211004296H}. Identifying massive WDs in young clusters is a requirement as it allows us to use the WD cooling age together with the cluster age to estimate the WD's progenitor mass. The breadth of modern stellar surveys greatly expands our ability to search for these objects; in particular, the precise parallaxes and proper motions measured by the Gaia survey \citep{2016A&A...595A...1G} allow us to select high-confidence cluster members using only astrometry and photometry. % particularly using the Milky Way stellar survey Gaia. 
%Unfortunately, despite the large amount of available data, 
Recently, a wide search for massive WDs in young clusters \citep{2021ApJ...912..165R} identified new young and high-mass WDs as cluster members, but failed to identify any WDs with masses in excess of $1.1~M_{\odot}$ or WDs with progenitors over 6.2~M$_{\odot}$, leaving a gap in the high-mass region of the WD initial-final mass relation.

\begin{figure*}[ht]
    \centering
    \includegraphics[width=0.5\textwidth]{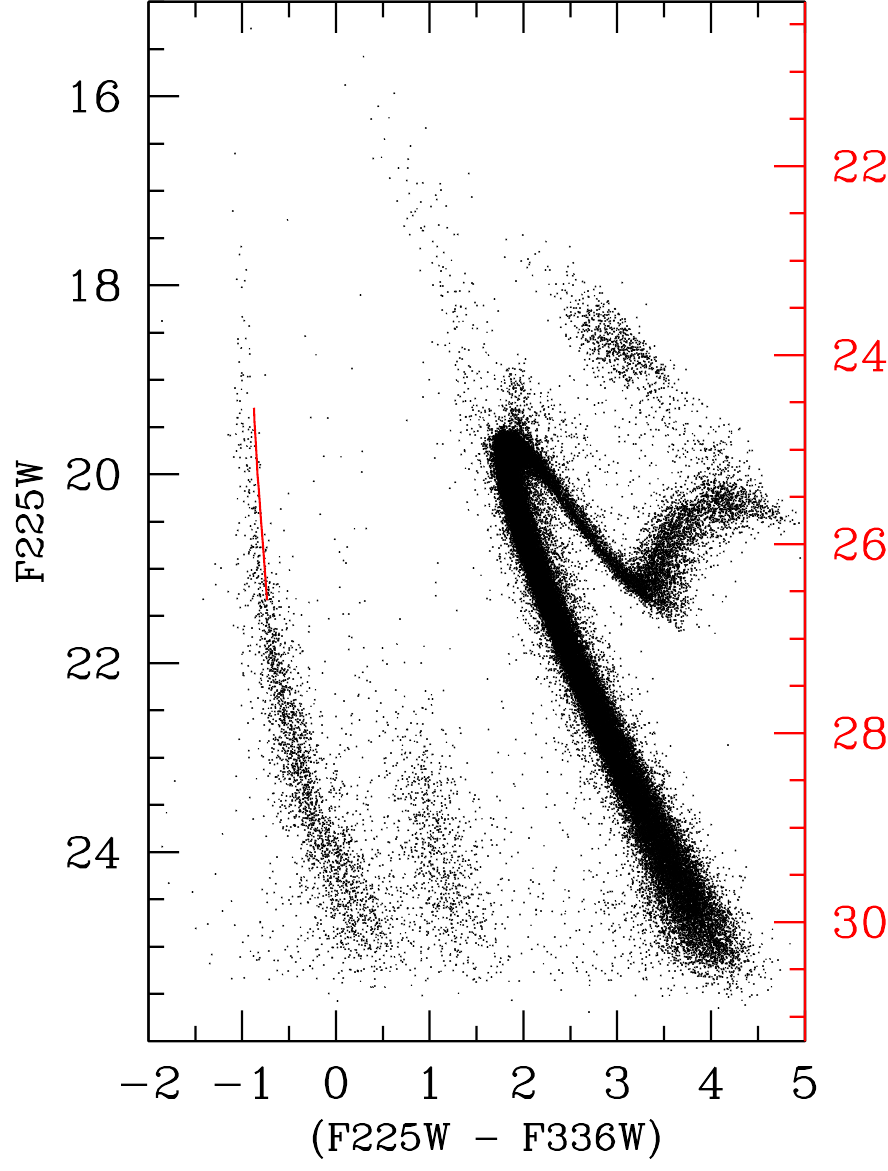}
        \caption{Observed colour-magnitude diagram of the globular cluster 47 Tucanae in the WFC3 filters F225W-F336W. 
        The left y-axis (apparent F225W magnitude) is for 47 Tuc itself while the right y-axis moves the cluster to the distance of the LMC. 
        The red line is a $1.0~M_{\odot}$ WD cooling sequence for WDs in the LMC for temperatures from 100,000~K - 30,000~K and illustrates that in these filters it should be possible to detect 
        hot WDs at the distance of the LMC. The feature between the 47 Tuc WDs and the main sequence of the cluster is the upper main sequence of the LMC.}
    \label{fig:cmd-47tuc}
\end{figure*}

As the most massive WDs are the first WDs produced in a cluster, and since escape velocities in open clusters are quite small, the missing massive WDs might have escaped their host clusters. Open clusters are in fact known to be deficient in WDs \citep{2003ApJ...595L..53F}, and this deficit is thought to occur from WDs receiving a natal velocity kick of about 1~km/s at birth \citep[see][and references therein]{2018MNRAS.480.4884E,2009ApJ...695L..20F,2007MNRAS.382..915H}. In order to increase the number of potential massive WD cluster members, we expanded our search to include WDs that may have escaped from their host clusters
\citep{2021arXiv211009668M, 2021arXiv211003837H}.

Even within this expanded search, the number of young massive clusters in the immediate solar vicinity is very small, with only 5 known (M~45 (Pleiades), IC~2602, IC~2391, IC~2451A, Alpha Persei) closer than 200 pc with ages less than 200 Myrs \citep{2021ApJ...912..165R}. Searching farther afield is likely to be unproductive as Gaia parallaxes become increasingly unreliable with larger distances and confusion with field WDs increases. A novel approach, which we exploit in this paper, is to search in the massive young clusters of the Magellanic Clouds, where contamination is greatly reduced and straight-forward observations can exclude Galactic contaminants. 

The clear issue is the faintness of a hot WD in the Magellanic Clouds. To illustrate that this is not insurmountable, we display the CMD of 47 Tuc in Fig.~\ref{fig:cmd-47tuc} \citep[from HST Program 12971 in Cycle 20, Richer PI, see][]{2016ApJ...821...27G}. The exposure times here were quite modest, with 1080 and 1205 seconds respectively in F225W and F336W in each of 9 different fields centered on the cluster. The well-defined WD cooling sequence is seen as the left-most feature in this plot. Hot WDs in these filters are amongst the brightest objects seen in such an old cluster. In a young Magellanic Cloud cluster this will not be the case as massive hydrogen-burning stars will still be present on the main sequence. 

The y-axis to the left is the appropriate one for 47 Tuc, while the right hand $y$-axis (in red) shifts the cluster to the distance of the LMC. The approximately vertical red line represents the cooling sequence for a $1.0~M_{\odot}$ WD with temperature between 100,000 and 30,000~K \citep{2020ApJ...901...93B,2011ApJ...730..128T}. At these very high temperatures,  massive WDs have not yet fully contracted and that is why they are as bright as the lower-mass WDs. These WDs will appear at an F225W magnitude of $\sim$24.5 in the LMC, which is reachable with HST with only modest exposure times. We see almost a full magnitude of the WD cooling sequence below this limit in this relatively short-exposure CMD.

\section{Magellanic Cloud Clusters}

In the following section we present the observed CMDs and other diagnostic diagrams of the four Magellanic Cloud young clusters in which we searched for WDs. For all these clusters we use mean distances and reddening values to both the LMC and SMC. The adopted true distance modulus to the LMC we employ is $18.48\pm0.03$ \citep{2019Natur.567..200P} while we use $18.98\pm0.03$ for the SMC \citep{2020ApJ...904...13G}. The reddening, E(B-V), in the directions of both galaxies was obtained from \cite{2019A&A...628A..51J}: 0.09 for the LMC and 0.04 for the SMC. The ratios of total to selective absorption we employed came from the STScI Users Guide: R$_V$ = 2.74 for the SMC and 3.41 for the LMC with A$_{F225W}$/A$_V$ = 2.63 and A$_{F336W}$/A$_V$ = 1.68 obtained from the online models of the Padova group (\url{http://stev.oapd.inaf.it}).

For each target cluster, individual \texttt{.flc} science images were retrieved from the HST archive, in addition to a deep stacked, drizzled, distortion-corrected \texttt{.drc} reference image.  We used the \texttt{Dolphot} package  \citep[][and later updates]{2000PASP..112.1383D} for subsequent preprocessing and 
point-spread function fitting (PSF) photometry.  Briefly, to prepare each image for photometry, \texttt{Dolphot} masks bad pixels and generates sky frames for each science image.  Next, \texttt{Dolphot} uses model PSFs customized to each filter of each instrument to iteratively perform simultaneous PSF photometry across all of the science images.  While \texttt{Dolphot} has numerous parameters governing the details how the PSF photometry is performed, we use the parameters recommended by the \texttt{Dolphot} manual with the exception that, building on previous experience \citep{2018ApJ...864..147C,2018AJ....156...41C}, we modify a few of these to the values used by \citet{2014ApJS..215....9W} to optimize photometry in the crowded inner regions of our target clusters\footnote{Specifically, we set \texttt{Force1}=1, which treats all sources as stellar, requiring judicious a posteriori quality cuts on our photometry to \\ eliminate spurious detections (see below).  In addition, we use the values of \citep{2014ApJS..215....9W} \texttt{RCombine}=1.415 and \texttt{PosStep}=0.1.}.

The magnitudes output by \texttt{Dolphot} have been placed on the Vegamag photometric system using photometric zeropoints and encircled energy corrections from   \citet{2017wfc..rept...14D} and \citet{2016wfc..rept....3D}. Bearing in mind that reliable color information is needed to select WD candidates, we employ several photometric quality cuts to eliminate spurious, non-stellar and/or unreliable detections from our photometric catalogs:

\begin{enumerate}

\item S/N$\geq$5 in both filters. See Fig. 3 (left) for what this means in terms of the WD cooling sequence in NGC 2164.
\item $|$\texttt{sharp}$|$$\leq$0.4. Sharp is 0 for a star that is perfectly fit by the PSF, positive for a star image that is too sharp (e.g. cosmic ray) and negative for an image that is too broad (galaxy, stellar blend). In uncrowded fields sharp will generally be between -0.3 and +0.3.
\item \texttt{crowd}$\leq$0.1. The crowding parameter is in magnitudes and tells how much brighter the star would have been if nearby stars had not been fit simultaneously. For an isolated star the value is 0.
\item A photometric quality flag $\leq$2 in each filter. A flag of 0 means that the star was recovered extremely well in the images. A value less than or equal to 2 means only keeping the stars that are not saturated,  are not extremely impacted by bad pixels, and are not affected by both an edge and bad pixels (which would sum to 3).  
\end{enumerate}

We performed artificial star tests to quantify incompleteness, photometric errors and bias (mean offset) in the color-magnitude region occupied by WD candidates.  To fully capture the dependence of observational effects on both luminosity and crowding, over 50,000 artificial stars were assigned input magnitudes 24$\leq$F225W$\leq$26, a fixed color of $($F225W$-$F336W$)$=$-$1, a flat luminosity function in F225W, and a spatial distribution following a realistic (i.e.~exponentially declining) radial cluster density profile.  Artificial stars are processed one at a time by \texttt{Dolphot}, so there is no limitation on the proximity of each artificial star to any other artificial star.

\begin{table*}
\caption{Magellanic Cloud Cluster Parameters}
\label{tab:clusters}
\centering
\begin{tabular}{lcccccc}
\hline
Cluster Name & $(m-M)_{0}$ & E(B-V) & $M_{V}$ (cluster) & [M/H] & Age (Myrs) & Progenitor Mass ($M_{\odot}$)\\
\hline
\hline
NGC 2164 & 18.48 & 0.09 & -8.4 & -0.42 & 60-100 & 6.5-5.1 \\  
NGC 1805 & 18.48 & 0.09 & -8.4 & -0.42 & 25-40 & 10.2-8.1 \\
NGC 330  & 18.98 & 0.04 & -9.7 & -0.94 & 32 & 9.0 \\
NGC 1818 & 18.48 & 0.09 & -9.8 & -0.42 & 25-40 & 10.2-8.1 \\
\hline
\end{tabular}
\end{table*}

\begin{table*}
\caption{Exposure Times (secs) of HST Images Used in WD Search}
\label{tab:exposure}
\centering
\begin{tabular}{lccccc}
\hline
Cluster  Name & F225W& F336W& F555W & F814W\\
\hline
\hline
NGC2164& 4800  & 3741 & 1400 & 848 \\ 
NGC1805 & 4800  & 3741 & .....& 756 \\
NGC330 & 4860 & 3795 & ..... & 770 \\
NGC1818 & 4800 & 3741 & ..... & 756 \\
\hline
\end{tabular}
\end{table*}

\subsection{NGC 2164}

NGC 2164 is likely the oldest of the clusters in our sample. In a recent paper \citep{2017ApJ...844..119L}, the cluster is dated to be between 50 and 100 Myrs old, while an earlier estimate \citep{2006ApJ...646..939M} places the cluster age closer to 80 Myrs. In order to choose the most appropriate age and simultaneously search for potential cluster WDs, we employed the UV data from our HST proposal \#13727 (PI Kalirai) with a 4800~sec exposure in F225W and 3741~sec in F336W. Additionally, shorter exposures from the MAST archives in F814W (848~sec, proposal \#14710) and F555W (700~sec, proposal \#8134, WFC2) were also used. These latter exposures were rather short and do not play a significant role in our searches. 

\begin{figure*}[ht]
    \centering
    \includegraphics[width=0.65\textwidth]{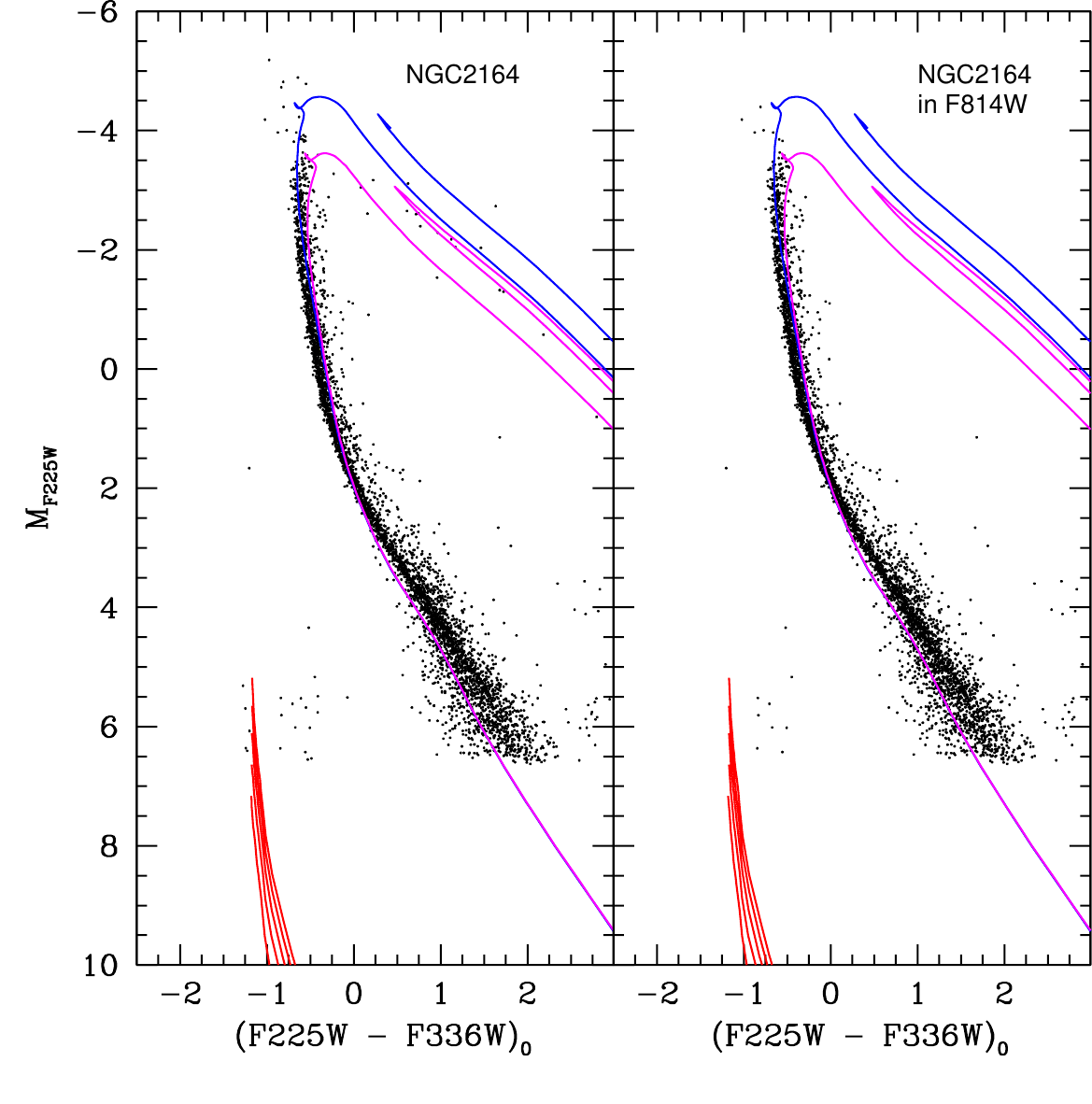}
   \caption{Colour-magnitude diagrams of the LMC cluster NGC 2164 in the WFC3 filters (F225W, F336W). In the left plot the Padova isochrones are for ages of 60 (blue) and 100 (cyan) Myrs and the CMD includes all stars that pass a mild error cut. The theoretical white dwarf cooling sequences (red) are for masses 0.9 - 1.3 in steps of $0.1~M_{\odot}$. The right plot, with the same isochrones and cooling models, additionally requires the stars to be present in our F814W images. There are no red giants here as they are saturated in the F814W exposures.}
    \label{fig:cmd-n2164}
\end{figure*}

The UV HST CMD of NGC 2164 is illustrated in Fig. 2 (left panel). Padova isochrones are shown for ages of 60 (blue) and 100 (cyan) Myrs, covering the range of earlier estimates. Assuming that most of the stars brighter than $M_{F225W} = -4$ are blue stragglers \citep{2017ApJ...844..119L}, an age between these limits (ie $\sim$80 Myrs) seems like the most appropriate. We adopt this age in the following discussion. At 80 Myrs old, and with the most massive current members (AGB stars) near 5.7~$M_{\odot}$, similar to some of the Milky Way clusters in which we have already located WD members \citep{2021ApJ...912..165R}, NGC 2164 is the most likely LMC cluster examined here to contain WDs.

The most remarkable feature in this colour-magnitude diagram is a putative cluster WD cooling sequence seen as the 5 objects in the WD region slightly blueward of the cooling models (in red). An additional potentially very luminous WD is seen 3 magnitudes brighter than these candidates. All the WD cooling models illustrated here terminate at a temperature of 150,000~K. 
Redward of these potential WDs, and well blue of the main sequence, there is a scatter of points. These are potentially WDs with red companions or background objects and simply for reference purposes we call these objects "Background Objects". Characterizing these as possibly binaries is a reasonable assumption as many appear in the right panel of  Fig.~\ref{fig:cmd-n2164}, which plots the same CMD but with the additional requirement that the objects are present in our F814W images as well. Cluster WDs that are single objects are not likely to be measurable in the rather short exposure F814W images as the apparent magnitude of a massive and hot WD ($\sim$80,000~K) in the LMC will be fainter than 29 in this filter, well below our detection limit. Taken together, this suggests that the objects near the cooling sequences are single WDs in the cluster (as they form a tight sequence) while the objects just to the red are potential WDs in binaries and may or may not be associated with NGC~2164 at all. Additionally, they may even be faint unresolved external galaxies many of which are known to have blue (F225W $-$ F336W) colours \citep[see for example][]{2013AJ....146..159T}.  

We can estimate how many young white dwarfs we expect to find from the field and from the cluster. Given the number of upper-main-sequence stars in the cluster, from the 80-Myr isochrone we expect about ten white dwarfs from the cluster to be within the observed field.  On the other hand, the turnoff location of the field stars in the (F225W-F555W) CMD is reasonably well represented by a 4-Gyr isochrone and, from the number of field sub-giants, we estimate that only a single non-cluster white dwarf should be detectable in the observed field.  \inserted{From both the 80~Myr cluster isochrone and a 4~Gyr isochrone representing the field, from the observed number of turnoff or giant stars we estimate the number of stars from each population that have recently completed their evolution to WDs, assuming a \citet{2001MNRAS.322..231K} initial-mass function.  We estimate that the WD birth rate from the cluster is about 6~Myr$^{-1}$ and from the non-cluster stars in the field 0.12~Myr$^{-1}$.  We assume that cluster WDs evolve as 1 Msun WDs which would be detectable for about 1~Myr, and for the non-cluster WDs we use the evolution of a 0.6 Msun WD  that would be detected for 9~Myr, resulting in the population estimates given above.}
%
% We can see a 0.6 SM WD brighter than M_336W 8.08 for 9e6 yr.
% We can see a 1.0 SM WD brighter than M_336W 8.08 for 1e6 yr.
% We can see a 1.1 SM WD brighter than M_336W 8.08 for 5e5 yr.
% The birthrate from the LMC is about 0.12 WD/Myr and from the cluster it is 5.5-6.6 WD/Myr.
% Therefore, we expect 5.6 WD from the cluster (1.0 SM) or half as many for 1.1 SM and 1 WD from the LMC.
%
\begin{figure*}
    \centering
\includegraphics[height=0.3\textwidth]{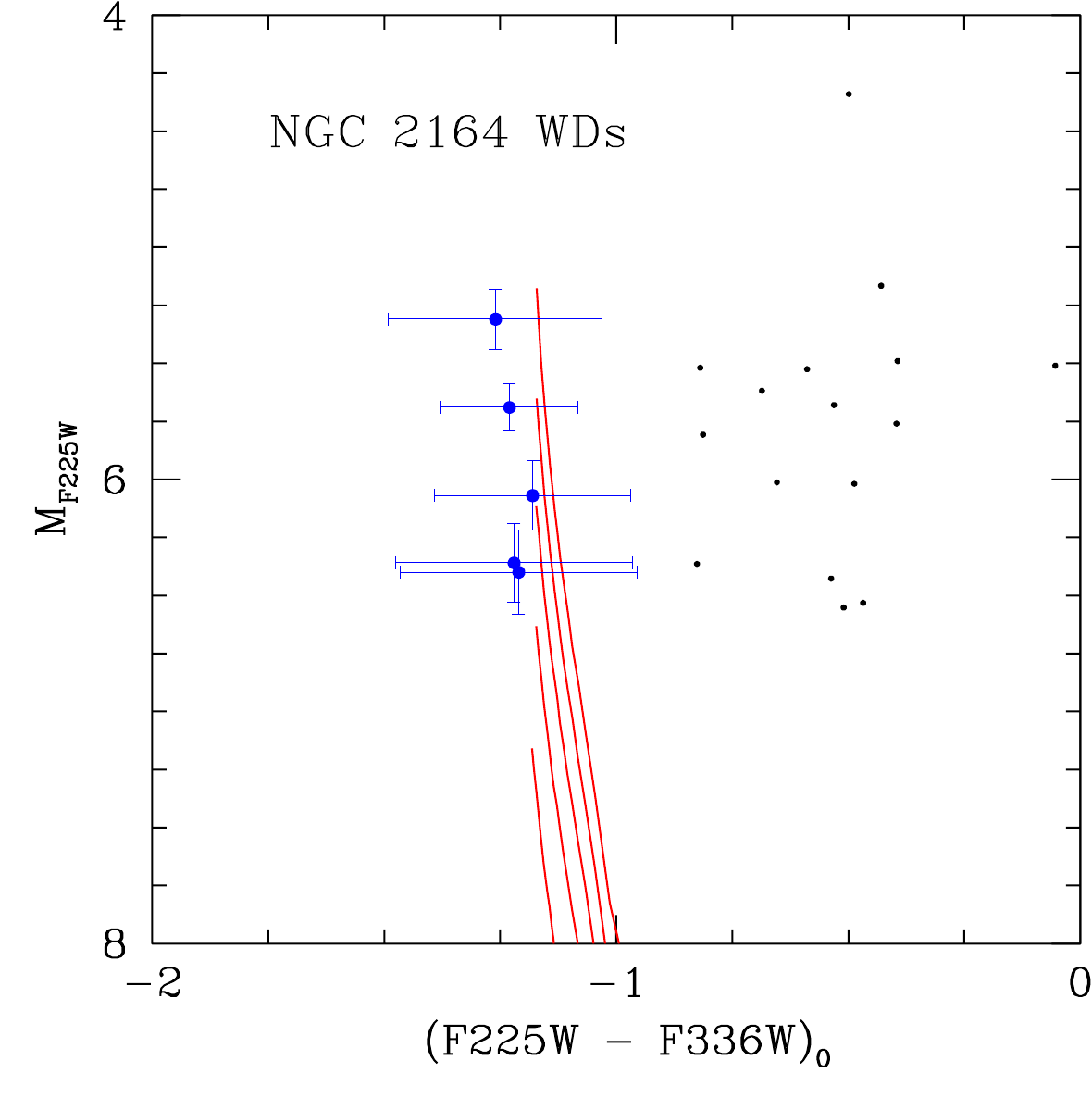}
\includegraphics[height=0.3\textwidth,trim=0 0.25in 0 0]{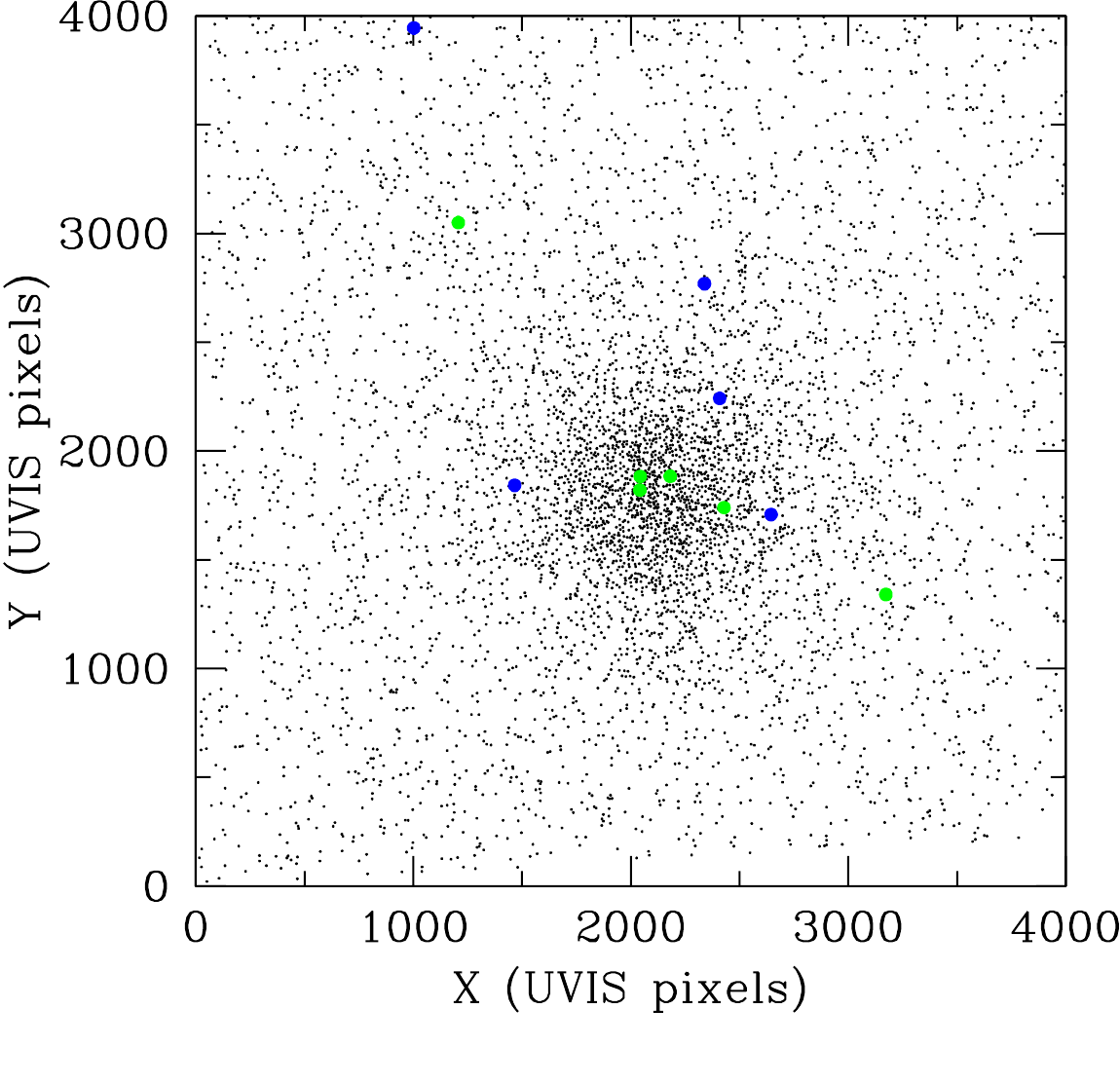}
\includegraphics[height=0.3\textwidth]{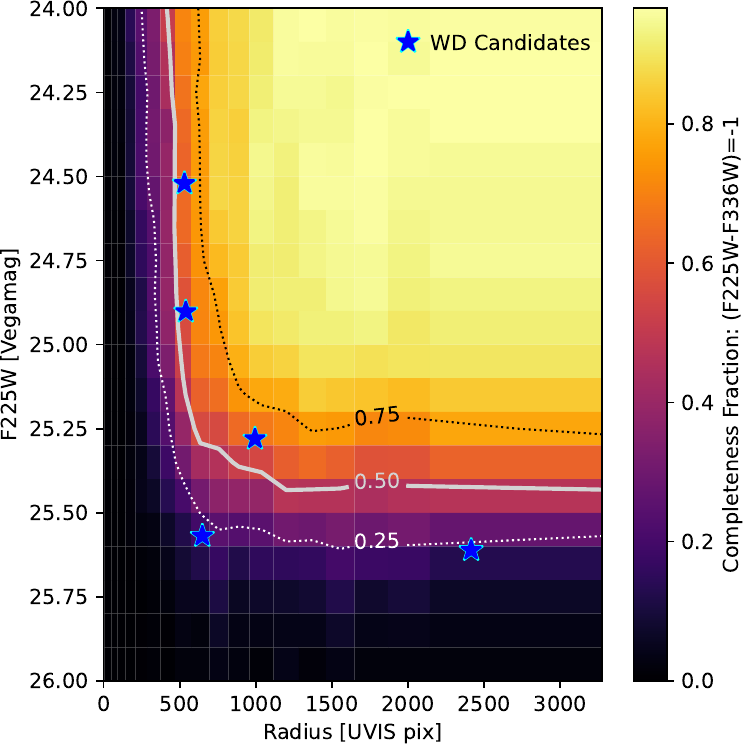}
         \caption{The diagrams in this figure explore the likelihood of the 5 potential NGC 2164 WDs being cluster members. The left plot is a blow up around the WD cooling sequence of NGC~2164 and includes error bars in the photometry for the WDs which are plotted in blue.  The middle plot indicates the position of the 5 potential WDs on the sky (blue points) with respect to the cluster as well as a selection of bright cluster stars (green). The rightmost diagram illustrates contour diagrams for the completeness calculations.}
    \label{fig:cc-n2164}
\end{figure*}

Fig.~\ref{fig:cc-n2164} provides further insight into the nature of these potential WDs. The left panel of this Figure is a blow up of the CMD around the WD cooling sequence and now includes the photometric errors for the individual stars. Within the error bars, all the putative WDs lie along the cooling sequences. The modest systematic offset toward bluer colours might be explained by a slight over-correction for the reddening in the UV filters for these very blue stars. 
The central plot of Fig. 3 illustrates the location of the NGC~2164 stars on the sky. In blue are indicated the positions of the 5 candidate WDs . In this plot, 1 pixel is 0.04 arcsecs corresponding to roughly 0.01 parsecs at the distance of the LMC. The potential WDs are obviously not highly concentrated to the cluster centre, but this, as we see below, is likely an incompleteness effect for these faint stars in a crowded cluster region. By contrast, in green, we indicate the locations of 6 bright blue stars near the tip of the main sequence.  These stars are significantly more centrally concentrated towards the cluster core than the WDs, suggesting that the WDs are either non-members, are too faint to be found in the crowded cluster core or have been ejected from the central regions of the cluster \citep{2021arXiv211009668M,  2021arXiv211003837H, 2021arXiv211004296H}. This latter scenario is unlikely, however, given the extremely short cooling times of these very hot WDs. 

The suggestion that the paucity of WDs towards the cluster centre is a selection effect due to crowding is borne out by incompleteness tests carried out on the images. Approximately 57,000 artificial stars were placed into the images with a fixed colour of (F225W~$-$~F336W)~$=-1$, 24$<$~F225W~$<$26 and a realistic power law cluster radial density profile. The right panel of Fig. 3 shows a 2-dimensional map of completeness as a function of F225W and radius in pixels from the cluster center.  The faintest candidates never get above 25\% completeness due to a lack of signal, whereas the brighter candidates are located at a radius where crowding starts to significantly impact completeness.  Taken together with the structural parameters of the cluster (half-light radius = 3.6 $-$ 3.7 arcsec = 90 $-$ 93 UVIS pixels for Wilson or exponential profiles and Wilson tidal radius $=$ 137.2 arcsec $=$ 3430 pixels \citep{2005ApJS..161..304M}), this implies that there are likely more WD candidates closer to the cluster center that are difficult to detect because of crowding. 

It is clear that all the putative WD cluster members are extremely hot, so hot in fact that we can make the reasonable assumption that their cooling ages are negligible in comparison to their lifetimes before becoming WDs. Taking this as the total evolutionary time of their progenitors up until the time they produce a WD (end of the AGB phase),  a star of metallicity [M/H] $=$ $-$0.42 with an evolutionary time to the end of the AGB of 80 Myrs has a mass of 5.7~$M_{\odot}$, establishing this as the progenitor mass of each of these potential WDs.

Table 3 presents the coordinates and properties of these 5 NGC~2164 WD candidates.

%\tablenotetext{1}{ Subramanian, A. and Sagar, R. 1995, A\&A, 297, 695}
%\tablenotetext{2}{Glatt, K. et al. 2010, A\&A 517, A50}
%\tablenotetext{3}{Li, C. et al. 2013, MNRAS 436, 1497}
%\tablenotetext{4}{Liu, Q. et al. 2009, A\&A 503, 469}
%\tablenotetext{5}{Keller, S.C. et al. 2000, AJ 119, 1748 }
%\tablenotetext{6}{ Siriani, M. et al. 2002, ApJ 579, 275. See also Li, et al. 2017, ApJ 844, 119. }

\begin{table*}
\caption{White Dwarf Candidates in the LMC Cluster NGC 2164}
\label{tab:wdinclusters}
\centering
\begin{tabular}{lcccccc}
\hline
WD Name & RA & Dec & F225W & F336W & M$_{F225W}$ & (F225W $-$ F336W)$_{0}$\\
\hline
\hline
NGC2164-WD1 &  89.747744183 &  -68.514278325 & 24.52($\pm$0.13) &  25.52($\pm$0.19) & 5.31 & -1.24 \\ 
NGC2164-WD2 & 89.734048835& -68.510266495& 24.90($\pm$0.10) & 25.87($\pm$0.11) & 5.69  & -1.21 \\  
NGC2164-WD3 & 89.762529274& -68.516498888& 25.28($\pm$0.15) & 26.20($\pm$0.15) & 6.07 & -1.16 \\  
NGC2164-WD4 & 89.728909790& -68.523185255& 25.57($\pm$0.17)  & 26.53($\pm$0.19) & 6.36  & -1.20 \\  
NGC2164-WD5& 89.786485192& -68.534011505& 25.61($\pm$0.18)& 26.56($\pm$0.18) & 6.40  & -1.19\\  
 \hline
\end{tabular}
\end{table*}

 \subsection{NGC 1805, NGC 330 and NGC 1818}
 
 The clusters NGC 1805 and 1818 are LMC clusters with literature ages near 40 million years  \citep{2017ApJ...844..119L}. From online Padova isochrones (\url{http://stev.oapd.inaf.it/cgi-bin/cmd}), clusters of this age and metallicity would have stars of 8.1~$M_{\odot}$ producing WDs today. It would be of some significance if either or both hosted WDs at this time in their history, making either the youngest cluster known to do so. The SMC young open cluster NGC 330 is potentially even younger at 32 Myrs implying an even more massive progenitor for any current WD cluster members.  
 
 \begin{figure*}
    \centering
    \includegraphics[width=0.75\textwidth]{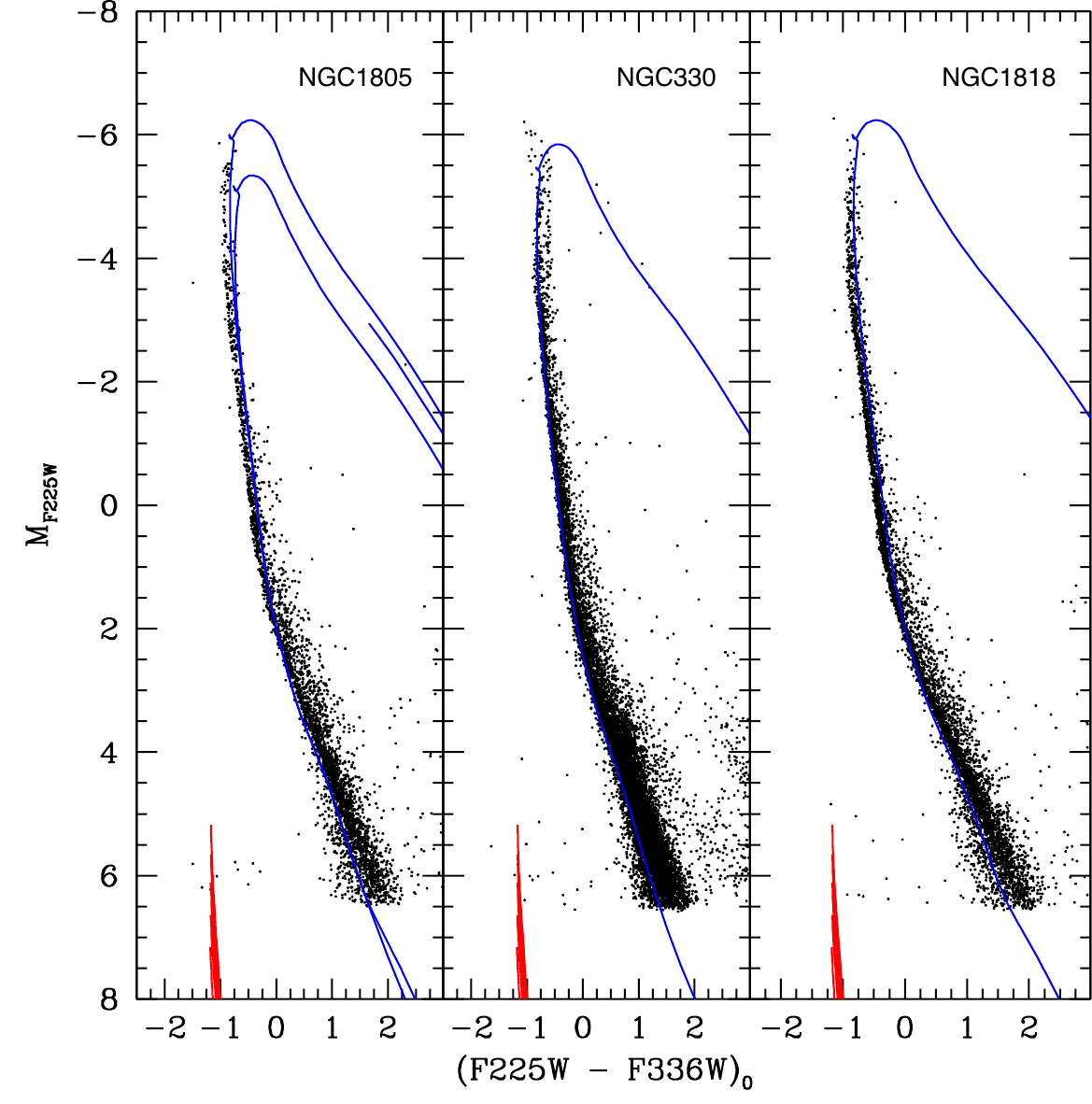}
     
         \caption{Colour-magnitude diagrams of the two young LMC clusters (NGC 1805, NGC 1818) and the SMC cluster NGC 330 in the WFC3 filters (F225W, F336W). Padova isochrones (blue) for 25 and 40 Myrs are shown for NGC 1805, 32 Myrs for NGC 330 and 25 Myrs for NGC 1818. The theoretical white dwarf cooling sequences (red) are for masses 0.9 - 1.3 in steps of $0.1~M_{\odot}$.}
    \label{fig:3panel.pdf}
\end{figure*}
 
 Fig. 4 plots the CMDs of these 3 clusters together with our best fitting isochrones for each. For NGC 1805 we used 25 and 40 Myr isochrones and found that the older age does not fit the CMD particularly well making our best age estimate for this cluster closer to 25 Myrs. The other LMC cluster (NGC 1818) was well represented by an isochrone of the same age, namely 25 Myrs. Stars just completing their AGB evolution at this age are $10.2~M_{\odot}$. NGC 330 appears to be around 32 Myrs and any member producing a WD today would have had an initial mass of $9.0~M_{\odot}$. These are the entries included in Table 1.
 
 In each cluster there is a scatter of stars near to or in the WD region of the CMD, but there is nothing resembling a cooling sequence as was found for NGC 2164. Colour-colour plots for these clusters were inconclusive as any third wavelength band was too short in exposure time to provide useful information.  
 
\subsection{Kolmogorov-Smirnov (K-S) Tests in Magellanic Cloud Clusters}
 
 \inserted{ To investigate the nature of these putative WDs, in Fig. 5 we explore the radial distributions using K-S tests of the various populations in NGC 2164 together with similar plots for the SMC cluster NGC 330 for comparison. Here, the various populations are as defined in the caption to Fig. 5. } 
 
 \begin{figure*}[ht]
    \centering
    \includegraphics[width=0.49\textwidth]{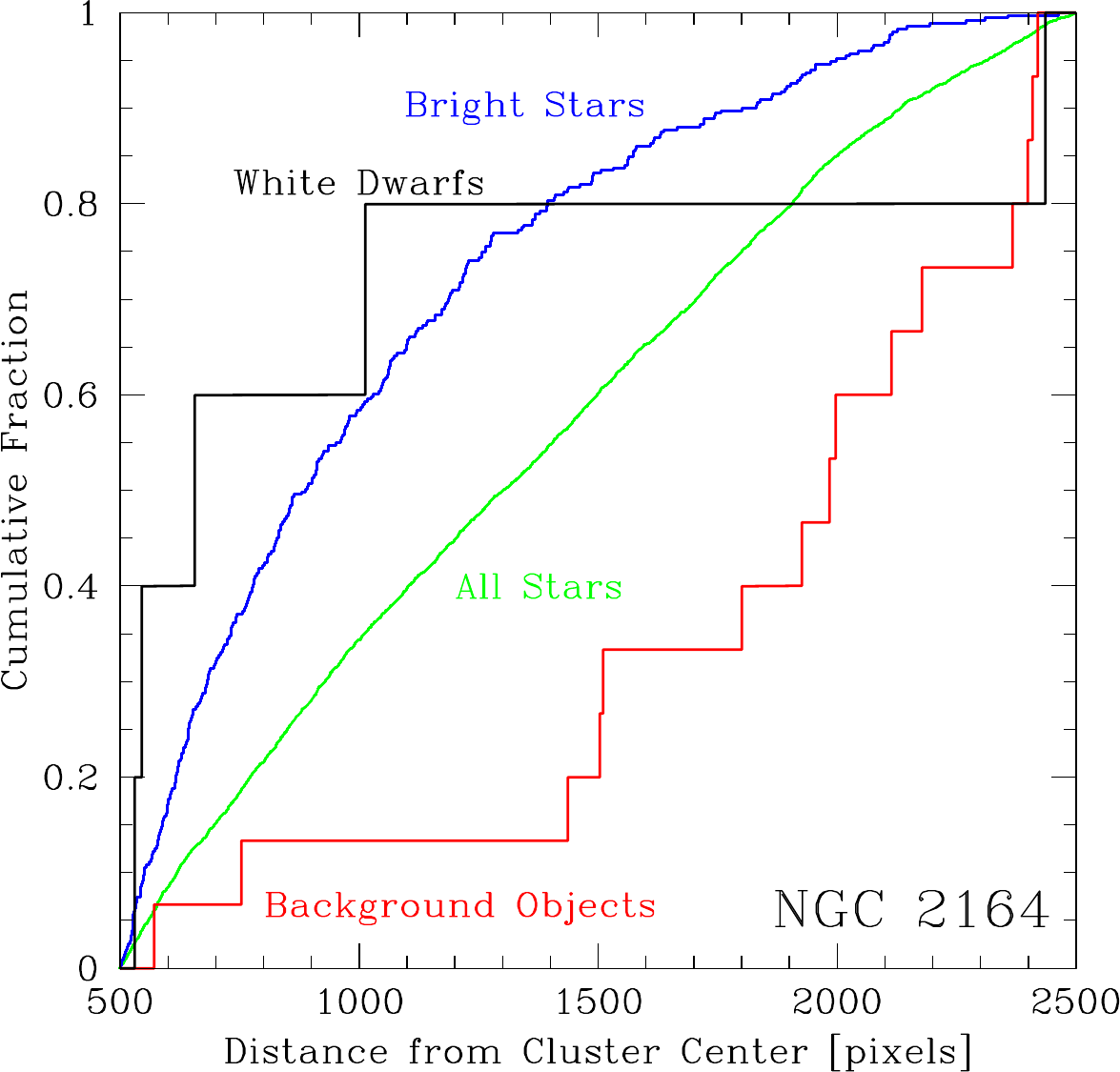} 
    \includegraphics[width=0.49\textwidth]{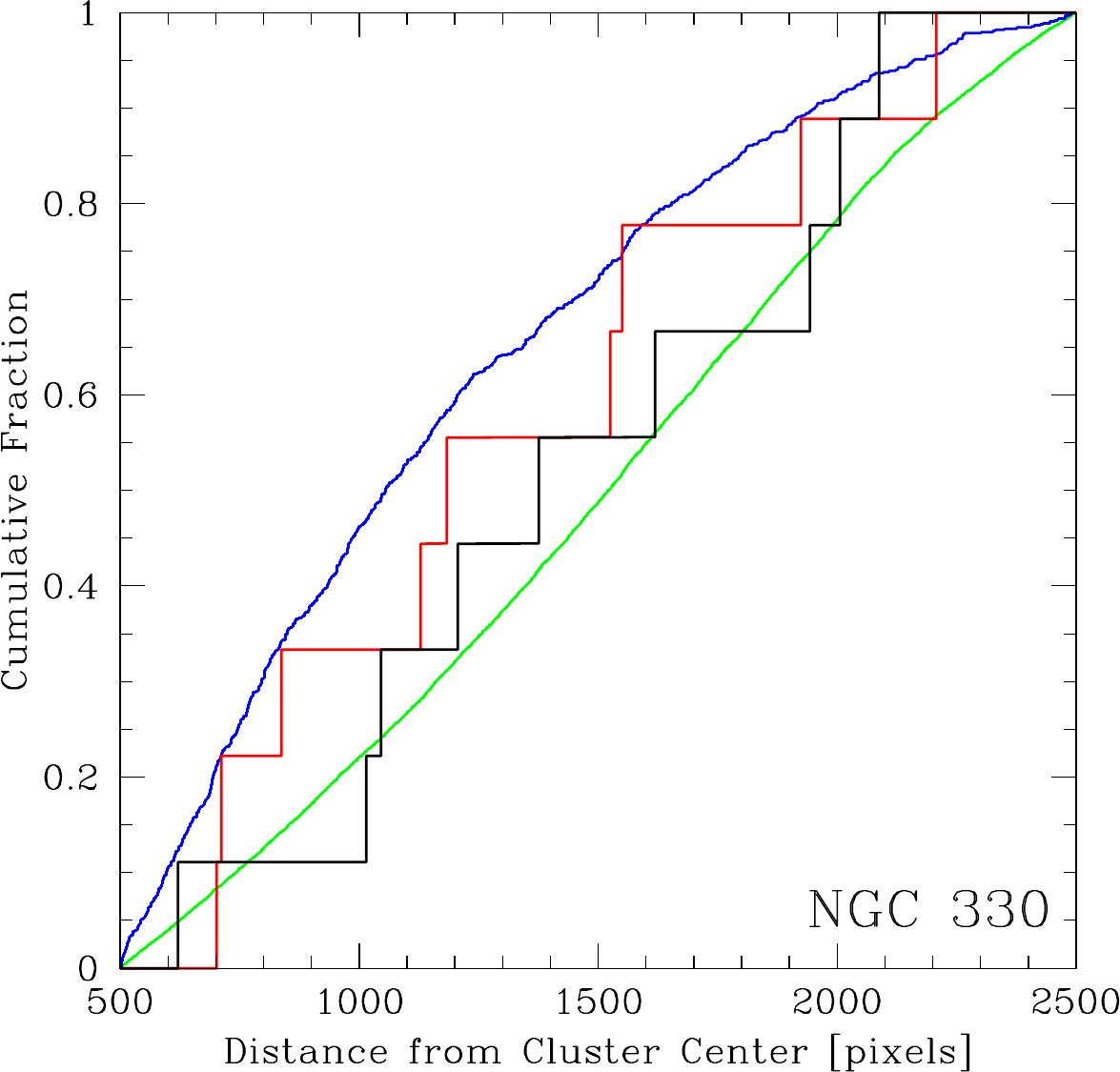} 
    \caption{Radial cumulative distribution for the different populations in the NGC 2164 and NGC 330 CMDs. In the left panel (NGC 2164), ``White Dwarfs'' consist of the 5 potential WD candidates, selected as absolute F225W magnitude fainter than +4 and colour bluer than $-1$. The ``Background Objects'' are the 15 objects seen just to the red of the putative cluster WDs in the CMD of Fig.~\ref{fig:cmd-n2164} with colours between $-1$ and 0. ``All Stars'' contains all stars in the sample and ``Bright Stars'' consists only of stars with absolute magnitude in F225W less than zero. The right panel shows the results of the same selections in the colour-magnitude diagram for NGC~330.}
    \label{fig:ks-n2164}
\end{figure*}

\inserted{ The 5 WDs in NGC 2164 have the appearance of being significantly more centrally concentrated than ``All Stars" in the cluster, but the small sample size yields a 13.6\% probability that the two were drawn from the same population. The ``Background Objects'' distribution certainly looks like a more extended population than ``All Stars'' and the K-S test verifies this returning only a 0.4\% chance that its population and that of all the cluster stars come from the same distribution. This is a significant result, with the suggestion that ``Background Objects'' may be unrelated to the cluster (background galaxies?). While the sample sizes are small for both the ``White Dwarfs'' and the ``Background Objects'', the K-S probability is only 3.8\% that they come from the same distribution. This is an important result as both samples are similarly faint, likely suffer similar incompleteness yet the ``White Dwarfs''  appear to be much more centrally concentrated.}  

\inserted {For NGC 300, all samples were adjusted for the different distance modulus and reddening of the SMC. The K-S plots here are less definitive. The putative WD sample (stars with absolute F225W magnitude less than +4 and (F225W - F336W) $< -1$ as defined for NGC 2164) does not exhibit a well defined cooling sequence as seen in NGC 2164 and neither is its radial distribution as dramatic. This cluster is both more distant and resides in a higher background than NGC 2164, both of which likely influenced these results.}

\inserted {Table 4 contains the full suite of K-S statistics for the four clusters. All categories were defined in the same way as for NGC 2164 (see caption to Figure 5) taking into account the different distance and reddening to the SMC. The main conclusion from this compilation and Fig. 5 are that the population of candidate WDs in NGC~2164 is much more centrally concentrated than in the other clusters shown by the difference in the distribution between the WD candidates and the Background Objects.}

 \begin{table*}
\caption{K-S Probabilities Various Populations in Four Magellanic Cloud Clusters}
\label{tab:clusters-ks}
\centering
\begin{tabular}{lcccccc}
\hline
Cluster Name & All-WD&Bright-WD& Background-WD&All-Background&All-Bright&\\
\hline
\hline
NGC 2164 & 0.136 & 0.499 & 0.038 & 0.004 & 0.000  \\
NGC 1805 & 0.445 & 0.853 &0.534 & 0.869 & 0.000 \\
NGC 330  & 0.810 & 0.544 & 0.833 & 0.289 & 0.000 \\
NGC 1818 & 0.288 & 0.084 & 0.665 & 0.856 & 0.000 \\
\hline
\end{tabular}
\end{table*}

\section{Conclusions}   
 The four Magellanic Cloud clusters examined ranged in age, by our estimates, from as young as 25 Myrs to as old as 80 Myrs. At the young end of this range, the stars in the AGB phase today in the LMC clusters NGC~1805 and NGC~1818 are as massive as 10.2~$M_{\odot}$. If these clusters were producing WDs, that would close the gap considerably in solving the issue of the apparent tension between the number of observed type II SNe and the upper limit to WD production \citep{2011ApJ...738..154H}. For neither of the clusters this young do we have any direct evidence from the current study that such massive main sequence stars are producing WDs. For both NGC 1805 and 1818 there is a small scatter of stars in the WD region of the CMD but no obvious sequence of cooling WDs. These objects are possibly cluster WDs with poor photometry, WDs in binaries, objects in the field unassociated with the cluster or extragalactic objects. The full resolution of these various options will likely have to wait until 30+ meter telescopes are available to carry out spectroscopy on these sources. These same comments are applicable to the $\sim$32 Myr SMC cluster NGC 330. For this cluster, any current WDs would have evolved from stars of mass $\sim$9.0~$M_{\odot}$. Only for the 80 Myr LMC cluster NGC 2164 was there evidence for a sequence of cooling WDs. Stars currently producing WDs in this cluster are $\sim$5.7~$M_{\odot}$, which is near the limit of what is currently being found within open clusters in the Milky Way \citep{2019ApJ...880...75R, 2020ApJ...901L..14C, 2021ApJ...912..165R} and even less than that derived for a few massive WDs that have escaped from their cluster environments \citep{2022ApJ...926L..24M,2021arXiv211004296H}.

Nevertheless, what does seem rather remarkable with this current study is that we have demonstrated that, with only modest exposure times in the UV with HST, it is possible to detect WDs in star clusters in nearby external galaxies.  

\section*{Acknowledgements}

Based on observations with the NASA/ESA {\it Hubble Space Telescope}, obtained at the Space Telescope Science Institute, which is operated by the Association of Universities for Research in Astronomy, Inc., under NASA contract NAS5-26555. 

This work was supported in part by NSERC Canada and Compute Canada via grants to H.R. and J.H.. I.C. is a Sherman Fairchild Fellow at Caltech and thanks the Burke Institute at Caltech for supporting her research. Support for this project was provided by NASA through grant HST-GO-13727 from the Space Telescope Science Institute, which is operated by the Association of Universities for Research in Astronomy, Inc., under NASA contract NAS5--26555.

\facilities{HST}.

%%%%%%%%%%%%%%%%%%%%%%%%%%%%%%%%%%%%%%%%%%%%%%%%%%



\bibliographystyle{aasjournal}
\bibliography{main} % if your bibtex file is called example.bib

%\begin{thebibliography}{99}
%\end{thebibliography}

%%%%%%%%%%%%%%%%%%%%%%%%%%%%%%%%%%%%%%%%%%%%%%%%%%

%%%%%%%%%%%%%%%%% APPENDICES %%%%%%%%%%%%%%%%%%%%%

\label{lastpage}

\end{document}